\documentclass{appolb}
\usepackage{epsfig}
\begin{document}
\title{Renormalization group flow equations from the 4PI equations of motion%
\thanks{Talk presented at the HIC for FAIR Workshop and XXXI Max Born Symposium
``Three days of critical behaviour in hot and dense QCD,'' Wroc\l aw, June 14-16, 2013.}%
}
\author{M.E. Carrington
\address{Department of Physics, Brandon University, Brandon, Manitoba, R7A 6A9 Canada}
}
\maketitle
\begin{abstract}
The 4PI effective action provides a a hierarchy of integral equations which have the form of Bethe-Salpeter equations. The vertex functions obtained from these equations can be used to truncate the exact renormalization group flow equations. This truncation has the property that the flow is a total derivative with respect to the flow parameter and is equivalent to solving the $n$PI equations of motion. This result establishes a direct connection between two non-perturbative methods.  
\end{abstract}
  
\section{Introduction}
\vspace*{-.1cm}

There is a lot of interest in the study of non-perturbative systems, which cannot be solved by exploiting the existence of a small expansion parameter. In this paper we discuss two formalisms that have been proposed to address non-perturbative problems: $n$-particle irreducible ($n$PI) effective theories, and  
the exact renormalization group (RG).  
The $n$PI formalism has been used to study finite temperature systems and
non-equilibrium dynamics (see \cite{Blaizot1999,Berges2001} and references therein), and transport coefficients \cite{Aarts2004,Eduard-all}.  The exact RG has been applied to a variety of problems \cite{Wetterich1993}. 

%It has been proposed that the hierarchy of RG flow equations could be truncated at the level of the first equation using the Bethe-Salpeter (BS) equation derived from the 2PI effective action \cite{Blaizot2011,Blaizot2010}. 
%The flow of the 2-point function is a total derivative with respect to the flow parameter, and the integral of the flow equation gives an integral equation whose solution is equivalent to the equation of motion (eom) for the 2-point function from the 2PI effective action. 
We show that the 4PI effective action produces two Bethe-Salpeter (BS) equations that can be used to truncate the RG flow equations at the level of the second equation, and that the resulting flow equations for the 2- and 4-point functions are total derivatives whose integrals give the 4PI eom's. This result is surprising (since the full hierarchy of  RG flow equations are obtained using a single bi-local source term), and suggests that a BS truncation at arbitrary orders produces equations whose integrals give the $n$PI eom's. 
%This establishes a direct connection between two non-perturbative methods. 
%It also means that the truncation of the RG equations at any level of the heirarchy can be systematically extended by adding more and more skeleton diagrams to the effective action. For the $n$PI formalism, there could be a practical advantage in reformulating the integral equations as flow equations, because initial value problems are usually easier to solve than non linear integral equations. 

%\section{Notation}
%\label{section:notation}
We work with a scalar field theory with quartic coupling and consider only the symmetric case where the expectation value of the field is zero. 
We use a compactified notation in which the space-time coordinates are represented by a single numerical subscript.  We also use an Einstein convention in which a repeated index implies an integration over space-time variables.

\vspace*{-.1cm}

\section{The $n$PI effective theory}
\label{section:nPI}

\vspace*{-.1cm}

\begin{figure}[H]
\begin{center}
\includegraphics[width=7cm]{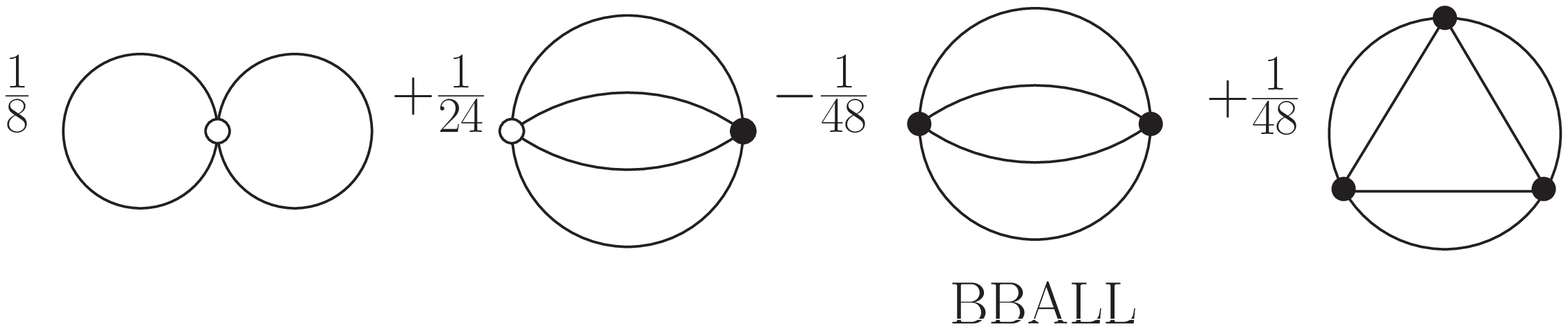}
\end{center}
\caption{The functional $\Phi_2$ for the 4-Loop 4PI effective action. 
%Bare vertices are denoted by open circles and effective vertices are solid dots. 
\label{fig:PhiAandB-2}}
\end{figure}

The $n$PI effective action is obtained by taking the $n$th Legendre transform of the generating functional which is constructed by coupling the field to $n$ source terms of the form $J_i\varphi_i+\frac{1}{2}R_{ij}\varphi_i\varphi_j+\cdots$. The effective action has the form
\begin{eqnarray}
\label{genericGamma}
\Gamma 
= S_{cl}[\phi]+
    \frac{i}{2} {\rm Tr} \,{\rm Ln}G^{-1}  +
\frac{i}{2} {\rm Tr}\left[ \left(G^0\right)^{-1} G\right] -i\Phi_2 \,,\nonumber
\end{eqnarray}
where $\Phi_2$ contains all contributions which have two or more loops. For example, $\Phi_2$ for the 4-Loop 4PI effective action in the symmetric theory \cite{Carrington2004} is shown in Fig. \ref{fig:PhiAandB-2}.
\par
The self consistent propagator and vertex are obtained through the variational principle by solving the equations produced by taking the functional derivative of the effective action and setting the result to zero. 
We define kernels:
\begin{eqnarray}
\label{LAM0nl}
&&\Lambda^{\rm disco}_{abcd\cdots rstu\cdots}+\Lambda_{abcd\cdots rstu\cdots}\\
&& = 2^{\# G} 4!^{\# V} \big(G^{-1}_{rr^\prime}\cdots \big)\frac{\delta^{(\# G)+(\# V)}\Phi}{\delta G_{ab}\delta G_{cd}\cdots\delta V_{r^\prime s^\prime t^\prime u^\prime}\cdots}\,.\nonumber
\end{eqnarray}
The factors ${\# G}$ and ${\# V}$ indicate the number of $G$'s and $V$'s with respect to which the functional derivative is taken.  
The inverse propagators truncate the legs that are left behind by the functional derivatives with respect to $V$, and they are moved through the derivative in (\ref{LAM0nl}) only so that the equation is easier to write - functional derivatives do not act on legs by definition. 
The definition of the kernel $\Lambda_{ab\cdots}$ also excludes a disconnected piece that contains only inverse propagators.

It is well known that the 2PI effective action gives a BS equation for a diagonal 4-point function.  
Writing the effective action as a functional of the form $\Phi[\phi,G,J(\phi,G),R(\phi,G)]$ we obtain: 
\begin{eqnarray}
\label{side1}
\frac{\delta}{\delta R_{ij}} \frac{\delta}{\delta G_{kl}}\Phi 
= -\frac{i}{4}\big(\delta_{ik}\delta_{jl}+\delta_{il}\delta_{jk}\big) = 
 \frac{\delta G_{xy}}{\delta R_{ij}}\frac{\delta^2\Phi}{\delta G_{xy}\delta G_{kl}}\,.
\end{eqnarray}
Calculating the derivative gives
\begin{eqnarray}
\label{GderR2}
&&- 2 i\frac{\delta G_{xy}}{\delta R_{ij}} =
 \langle \varphi_x \varphi_y \varphi_i \varphi_j\rangle-\langle \varphi_x\varphi_i\rangle  \langle \varphi_y\varphi_j\rangle -\langle \varphi_x\varphi_j\rangle  \langle \varphi_y\varphi_i\rangle \nonumber\\
&&=G_{ia}G_{jb}G_{xc}G_{yd}M_{abcd}+G_{ix}G_{jy} +G_{iy}G_{jx} \,.
\end{eqnarray}
Substituting (\ref{GderR2}) into (\ref{side1}) and using the kernel definition (\ref{LAM0nl}) produces:
\begin{eqnarray}
\label{BSfirst-coord}
M_{xykl}=\Lambda_{xykl}+\frac{1}{2} \Lambda_{xyab}G_{ac}G_{bd} M_{cdkl}\,.
\end{eqnarray}

\begin{figure}[H]
\begin{center}
\includegraphics[width=14cm]{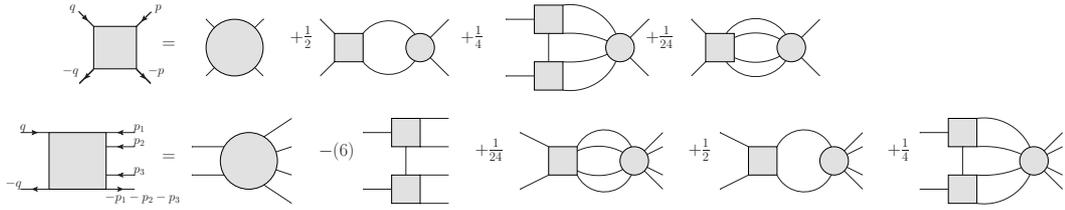}
\end{center}
\caption{BS equations from the 4PI effective action.\label{fig:BS-combo}}
\end{figure}

Higher order BS equations are calculated the same way \cite{Carrington2012b,Carrington2013}. Using the 4PI effective action (\ref{side1})  becomes:
\begin{eqnarray}
\label{side2-4PI}
% \frac{\delta^2 \Phi}{\delta R_{ij}\delta G_{kl}}=
\frac{\delta}{\delta R_{ij}} \frac{\delta}{\delta G_{kl}}\Phi  = 
-\frac{i}{4}\big(\delta_{ik}\delta_{jl}+\delta_{il}\delta_{jk}\big)
=  \frac{\delta G_{xy}}{\delta R_{ij}}\frac{\delta^2\Phi}{\delta G_{xy}\delta G_{kl}} +  \frac{\delta V_{xywz}}{\delta R_{ij}}\frac{\delta^2\Phi}{\delta V_{xywz}\delta G_{kl}}\,.\nonumber
\end{eqnarray}
The method used to obtain (\ref{GderR2}) gives:
\begin{eqnarray}
\label{VderR2}
\frac{\delta V_{xywz}}{\delta R_{ab}}  = \frac{i}{2}G_{a s} G_{b t}\bigg[V_{w x y z st} + (6) G_{uv}  V_{x z t u} V_{w y s v}\bigg]\,,\nonumber
\end{eqnarray}
and combining pieces we obtain the result in the first line of Fig. \ref{fig:BS-combo}. 
Similarly we obtain a BS equation for a  6-point function from $\delta\Phi/\delta R\delta V$:
\begin{eqnarray}
\label{BS24-a}
&&\frac{\delta^2 \Phi}{\delta R_{ab}\delta V_{cdef}}  = 
\frac{\delta G_{xy}}{\delta R_{ab}}\;\frac{\delta^2\Phi}{\delta G_{xy}\delta V_{cdef}} + \frac{\delta V_{xywz}}{\delta R_{ab}} \; \frac{\delta^2\Phi}{\delta V_{xywz}\delta V_{cdef}}\,.\nonumber
\end{eqnarray}
Substituting the derivatives and kernels gives, after a long calculation, the second line in Fig. \ref{fig:BS-combo} \cite{Carrington2012b}.

\vspace*{-.1cm}

\section{Renormalization group flow equations}
\label{section:RGflow}

\vspace*{-.1cm}

\begin{figure}[H]
\begin{center}
\includegraphics[width=10cm]{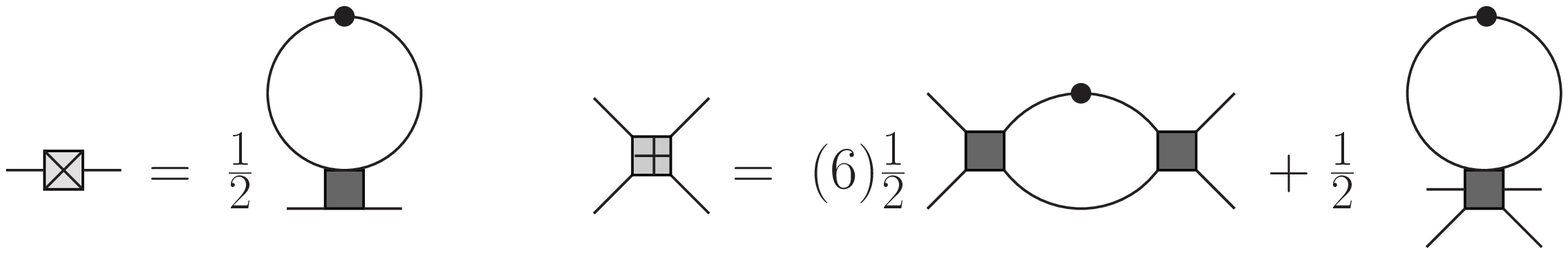}
\end{center}
\caption{Representation of equations (\ref{flow1}) and (\ref{flow2}). 
%Boxes with crosses  represent the derivative of the vertex  with respect to $\kappa$. The solid dot on a propagator is the insertion $\partial_\kappa R_\kappa$.  
\label{fig:renormGroup}  }
\end{figure}

The RG is constructed by building a family of theories indexed by a continuous parameter $\kappa$, with the dimension of a
momentum, such that fluctuations are smoothly taken into account as $\kappa$ is lowered from
the microscopic scale $\Lambda$ (at which the couplings are defined) down to zero. We add to the original action a non-local term which is quadratic in the fields $(\int_Q = \int d^4q/(2\pi)^4)$:
\begin{eqnarray}
\Delta S_{\kappa} = \frac{1}{2}\int_Q\, {\cal R}_\kappa(q)\varphi(q)\varphi(-q)\,.
\end{eqnarray}
The function ${\cal R}_\kappa(q)$ approaches zero for $q \ge  \kappa$, so that modes $\varphi(q \ge \kappa)$ are unaffected, and $\kappa^2$ for $q\ll \kappa$, which suppresses modes $\varphi(q \ll  \kappa)$  by giving them a mass $\sim\kappa$.
Generating functionals are defined in the usual way but with the action $S$ replaced by $S+\Delta S_\kappa$. We obtain ($\Phi=i\Gamma$, $R_\kappa = i {\cal R}_\kappa$): 
\begin{eqnarray}
\label{Gkappa-def}
&& \partial_\kappa \Phi_\kappa=  \frac{1}{2}\int_Q \partial_\kappa R_\kappa(q)  G^\kappa(q,-q)\,;~~~-(G^\kappa)^{-1} = R_\kappa + \Phi_\kappa^{(2)}\,.
\end{eqnarray}
Functionally differentiating (\ref{Gkappa-def}) with respect to $\phi$ produces the exact RG flow equations which form an infinite coupled hierarchy. The first two flow equations are (see Fig. \ref{fig:renormGroup}):
\begin{eqnarray}
\label{flow1}
\!\!\!\!\!\!&&\partial_\kappa \Phi_\kappa^{(2)} = \frac{1}{2}\int_Q \,(G_\kappa\partial_\kappa R_\kappa G_\kappa) \Phi_\kappa^{(4)}\,, \\
\label{flow2}
\!\!\!\!\!\!&&\partial_\kappa \Phi_\kappa^{(4)} = (6)\frac{1}{2}\int_Q\, \Phi_\kappa^{(4)} (G_\kappa\partial_\kappa R_\kappa G_\kappa)\,G^\kappa\Phi_\kappa^{(4)}
+ \frac{1}{2}\int_Q \,(G_\kappa\partial_\kappa R_\kappa G_\kappa) \Phi_\kappa^{(6)}\,.
\end{eqnarray}
The factor (6) in Eq. (\ref{flow2}) and Fig. \ref{fig:renormGroup} is a short-hand notation which means that there are 6 permutations of external legs. We use numerical factors in brackets in equations (figures) to represent additional terms that correspond to permutations of external indices that are not written (drawn). 
\par
\begin{figure}[H]
\begin{center}
\includegraphics[width=6cm]{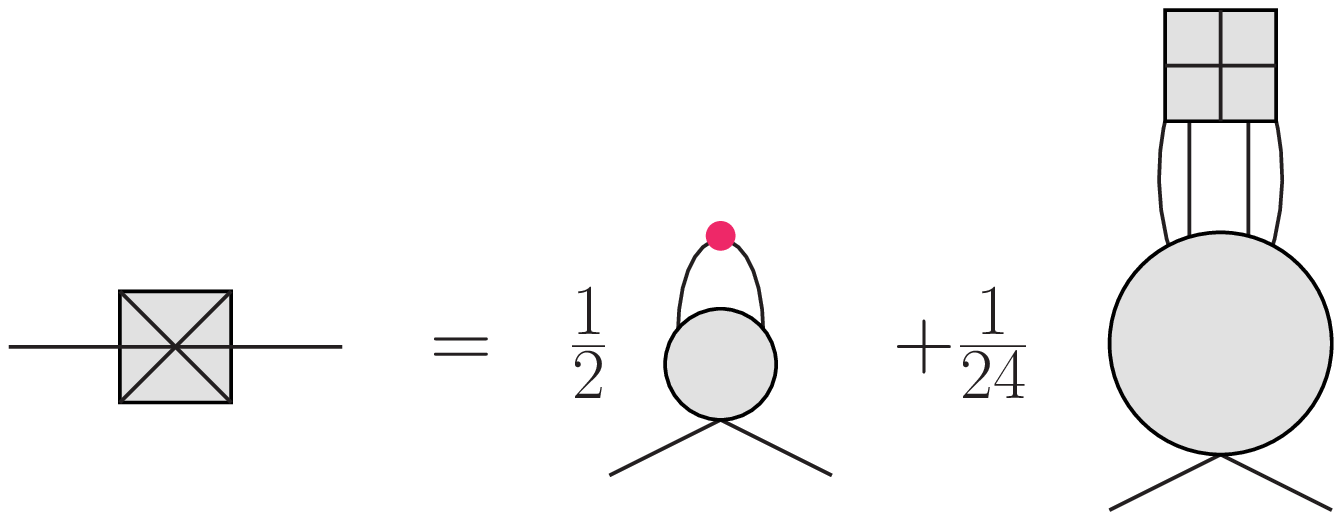}
\end{center}
\caption{The diagrams produced when the BS equation in the first line of Fig. \ref{fig:BS-combo} is substituted into the tadpole diagram in the first part of Fig. \ref{fig:renormGroup}.  \label{fig:22truncation-4PI}}
\end{figure}

\begin{figure}
\begin{center}
\includegraphics[width=8cm]{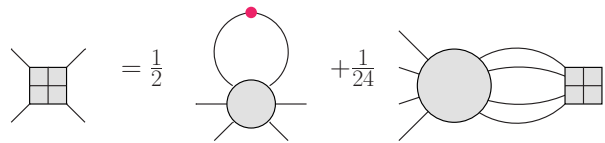}
\end{center}
\caption{
The diagrams produced when the BS equation in the second line of Fig. \ref{fig:BS-combo} is substituted into the tadpole diagrams in the second part Fig. \ref{fig:renormGroup}. \label{fig:BS24truncation-all}}
\end{figure}

 
In order to do calculations, the infinite hierarchy of RG flow equations must be truncated. This is a common feature of non-perturbative methods, and often leads to difficulties (see for example \cite{Smit2003}). 
%%%% In this paper we explore the relationship between the RG flow equations and the integral equations produced by the $n$PI formalism. 
The hierarchy of RG flow equations can be truncated at the level of the first equation using the Bethe-Salpeter (BS) equation derived from the 2PI effective action \cite{Blaizot2011}. 
%The RG flow equation for the vertex $\Phi^{(n)}$ is coupled to the vertex $\Phi^{(n+2)}$ through the tadpole diagram. 
%It is clear that the BS equation obtained from the derivative $\delta^2 \Phi_{n{\rm PI}}/\delta R_{ab}V_{xyzwu\cdots}$  will have two tadpole-legs with momenta $\pm q$, and therefore it is natural to truncate the hierarchy of RG equations using this equation. 
We can perform the truncation at the 4PI level using the BS equations in Fig. \ref{fig:BS-combo}. We extend these equations to the deformed theory by using kernels obtained from functional derivatives of the 4PI effective action which are subsequently evaluated at $G=G^\kappa$. 
Diagrammatically, we substitute the BS equations in Fig. \ref{fig:BS-combo} into the tadpole diagrams in Fig. \ref{fig:renormGroup}. After a long calculation  this produces the diagrams in Figs. \ref{fig:22truncation-4PI} and \ref{fig:BS24truncation-all} \cite{Carrington2012b}. We  represent the sum of terms $\partial_\kappa R_\kappa+\partial_\kappa \Phi^{(2)}$ by a small grey dot (red on-line).  

We compare our results for $\partial_\kappa\Phi_\kappa^{(2)}$ and $\partial_\kappa\Phi_\kappa^{(4)}$ with the derivatives $\partial_\kappa\Sigma_\kappa$ and $\partial_\kappa V_\kappa$  %of the 2- and 4-point functions 
using the 4PI equations of motion extended to the deformed theory. 
%Note that the vertex $V_\kappa$ depends on the flow parameter $\kappa$ only indirectly through the fact that the equation of motion for the 4-point function is coupled to the equation of motion for the 2-point function, which is evaluated at $G=G_\kappa$ after the functional derivatives are taken.
%In the 4PI theory equation (\ref{sigma-2PI-b}) becomes:
\begin{eqnarray}
\label{sigma-4PI-b}
&& \partial_\kappa(\Sigma_{ij})_\kappa = 2\partial_\kappa (G_{kl})_\kappa \frac{\delta^2\Phi}{\delta G_{kl}\delta G_{ij}}\bigg|_{G_\kappa} \\
&&+
2\partial_\kappa (V_{klrs})_\kappa (G_{ka})_\kappa (G_{lb})_\kappa  (G_{rc})_\kappa  (G_{sd})_\kappa  \;G^{-1}_{aa^\prime}G^{-1}_{bb^\prime}G^{-1}_{cc^\prime}G^{-1}_{dd^\prime}  \frac{\delta^2\Phi}{\delta V_{a^\prime b^\prime c^\prime d^\prime}\delta G_{ij}}\bigg|_{G_\kappa} \nonumber\\
&& = \frac{1}{2}\partial_\kappa (G_{kl})_\kappa (\Lambda_{ijkl})_\kappa + \frac{1}{24}\,\partial_\kappa (V_{klrs})_\kappa (G_{ka})_\kappa (G_{lb})_\kappa  (G_{rc})_\kappa  (G_{sd})_\kappa  (\Lambda_\kappa)_{abcdij}\,.\nonumber
\end{eqnarray}
Equation (\ref{sigma-4PI-b}) is precisely the result that is shown in Fig. \ref{fig:22truncation-4PI} if we identify $V_\kappa = \Phi^{(4)}_\kappa$, which means we have
\begin{eqnarray} \label{2result}
\partial_\kappa \Phi_\kappa^{(2)}(p) = \partial_\kappa \Sigma_\kappa(p)\,.
\end{eqnarray}
For the 4-point function a similar calculation gives \cite{Carrington2012b}.
\begin{eqnarray}
\label{4result}
\partial_\kappa \Phi_\kappa^{(4)}(p_1,p_2,p_3,-p_1-p_2-p_3) = \partial_\kappa V_\kappa(p_1,p_2,p_3,-p_1-p_2-p_3)\,.
\end{eqnarray}
Equations (\ref{2result}) and (\ref{4result}) show that the RG equations for the 2- and 4-point functions are total derivatives whose integrals can be written as the  4PI equations of motion.

\vspace*{-.1cm}
 
\section{Conclusions}
\label{section:conclusions}

\vspace*{-.1cm}

We have shown that the BS truncation produces flow equations that are total derivatives with respect to the flow parameter, and that the integrals of the flow equations give the equations of motion of the 4PI effective theory. 
This establishes a direct connection between two non-perturbative methods. 
%It also means that the truncation of the RG equations at any level of the heirarchy can be systematically extended by adding more and more skeleton diagrams to the effective action. 
For the $n$PI formalism, there could be a practical
advantage in reformulating the integral equations as flow equations, because initial value
problems are usually easier to solve than non linear integral equations. 
Furthermore, regarding the vertices from the $n$PI effective theory as flow equations gives new insight into the problem of how to renormalize the $n$PI effective theory for $n>2$.


\vspace*{-.1cm}


\begin{thebibliography}{}
%\bibitem{Jackiw1974} J.M. Cornwall, R. Jackiw, and E. Tomboulis, Phys. Rev. {\bf D 10}, 2428 (1974).
%\bibitem{Norton1975} R.E. Norton and J.M. Cornwall, Annals of Physics {\bf 91}, 106  (1975).

\bibitem{Blaizot1999} J.~P.~Blaizot, E.~Iancu, and A.~Rebhan, Phys. Rev. Lett. {\bf 83}, 2906 (1999); Phys. Rev. {\bf D 63}, 065003 (2001); J.~Berges, Sz.~Bors\'{a}nyi, U.~Reinosa, and J.~Serreau, Phys. Rev.{\bf D 71}, 105004 (2005).
%
%
\bibitem{Berges2001} J.~Berges and J.~Cox, Phys. Lett. {\bf B 517}, 369 (2001); J.~Berges, Nucl. Phys. {\bf A 699}, 847 (2002); G.~Aarts and J.~Berges, Phys. Rev. Lett. {\bf 88}, 041603 (2002); G.~Aarts, D.~Ahrensmeier, R.~Baier, J.~Berges, and J.~Serreau, Phys. Rev. {\bf D 66}, 045008 (2002).
%
\bibitem{Aarts2004} G.~Aarts and J.~M.~Mart\'{i}nez Resco, JHEP {\bf 02}, 061 (2004). 
\bibitem{Eduard-all} M.E. Carrington and E. Kovalchuk,  Phys.Rev. {\bf D77},  025015 (2008); ibid, Phys. Rev. {\bf D80}, 085013 (2009); ibid, Phys. Rev. {\bf D81}, 065017 (2010).
%
\bibitem{Wetterich1993} C. Wetterich, Phys. Lett., {\bf B301}, 90 (1993); U. Ellwanger, Z. Phys. {\bf C58}, 619 (1993);  N. Tetradis and C. Wetterich, Nucl. Phys. {\bf B422}, 541 (1994); T.R. Morris, Phys. Lett. {\bf B329}, 241 (1994);  J. Berges, N. Tetradis and C. Wetterich,  Phys. Rept. {\bf 363}, 223 (2002);  J. M. Pawlowski, Annals Phys. {\bf 322} 2831, (2007).
%
%
\bibitem{Carrington2004} M.E. Carrington, Eur. Phys. J. {\bf C35}, 83 (2004); J.~Berges, Phys. Rev. {\bf D70}, 105010 (2004).
%
%
\bibitem{Carrington2012b} M.E. Carrington, Phys. Rev. {\bf D87}, 045011 (2013)
\bibitem{Carrington2013} M.E. Carrington, WeiJie Fu, T. Fugleberg, D. Pickering, I. Russell,  Phys. Rev. {\bf D88}, 085024 (2013).
%
\bibitem{Smit2003} A. Arrizabalaga and J. Smit, Phys. Rev. {\bf D66}, 065014 (2002);  M.E. Carrington, G. Kunstatter and H. Zaraket, Eur. Phys. J. {\bf C 42}, 253 (2005); D. Binosi and J. Papavassiliou, Physics Reports, {\bf  479}, 1 (2009).
%
%\bibitem{Morris-98} T. R. Morris, Prog. Theor. Phys. Suppl. {\bf 131}, 395 (1998); Int. J. Mod. Phys. {\bf B12}, 1343 (1998);
%\bibitem{Bagnuls2001}  C. Bagnuls and C. Bervillier, Phys. Rept. {\bf 348}, 91 (2001).

%\bibitem{Delamotte2007}  B. Delamotte, cond-mat/0702365.
%\bibitem{Rosten2007} O. J. Rosten, arXiv:1003.1366.
%
\bibitem{Blaizot2011} J.P. Blaizot, Phil. Trans. R. Soc. {\bf A 369}, 2735 (2011); J.P. Blaizot, J.M. Pawlowski, U. Reinosa, Phys. Lett. {\bf B 696}, 523 (2011).



%

%
%\bibitem{jaxo} D. Binosi and L. Theussl, Comput. Phys. Commun. {\bf 161}, 76 (2004).
%

%
%\bibitem{vanHees2002} H. van Hees and J. Knoll, Phys. Rev. D 65, 025010 (2002); ibid. 105005 (2002).
%

%\bibitem{Serreau2010} U. Reinosa and J. Serreau, Annals Phys. 325, 969 (2010).


\end{thebibliography}
\end{document}